\documentstyle[11pt,newpasp,twoside,epsf]{article}
\def\edcomment#1{\iffalse\marginpar{\raggedright\sl#1\/}\else\relax\fi}
\reversemarginpar

\begin{document}
\title{The Relativistic Binary Pulsar B1913+16}
 \author{Joel M. Weisberg}
\affil{Department of Physics and Astronomy, Carleton College, 
Northfield, MN, 55057 USA}
\author{Joseph H. Taylor}
\affil{Department of Physics, Princeton University,
Princeton, NJ 08544 USA}

\begin{abstract}We describe the results of a relativistic analysis of our observations of binary pulsar B1913+16, up to the latest 
measurements in 2001 August.\end{abstract}

\section{Introduction}

The relativistic binary pulsar B1913+16, discovered by Hulse \& Taylor
(1975), has now
been observed for twenty--seven years.  After a several--year hiatus due
to the Arecibo upgrading project, we have resumed regular timing and
profile measurements.  In this paper, we describe the results of our relativistic analysis of these data.

\section{Timing and Physical Parameter Extraction}

In order to optimize timing data acquisition, we and colleagues have 
developed a 
number of specialized backends  for recording binary and millisecond pulsar data, including a sweeping local oscillator system at 430 MHz (McCulloch, Taylor, \& Weisberg 1979) and a succession of wideband samplers and
signal averagers for use at 21 cm (Taylor \& Weisberg 1982; Rawley 1986; Stinebring et al 1992).
Most of our highest quality timing data were acquired with the Princeton
Mark III backend (Stinebring et al 1992), which typically achieved time of arrival uncertainties $\Delta$TOA $\sim16 \ \mu$s in five--minute integrations 
of 40 MHz total bandwidth.  

We have found that the quality of data acquired
starting in 1981 are so much higher than earlier data that the previous 
measurements contribute negligibly to our fits.  Conseqently, we use 5083
TOAs measured between 1981 and 2001 in what follows.
These TOAs were fed into program TEMPO (Taylor \& Weisberg 1989; 
http://pulsar.princeton.edu/tempo/) which fitted
for 18 parameters according to the phenomenological relativistic timing 
model of Damour \& Deruelle (1985, 1986).

Aside from pulsar spin and astrometric parameters, we fitted for five ``Keplerian'' parameters and three relativistic quantities:  projected 
semimajor axis of the pulsar orbit $a_p \sin i$, 
eccentricity $e$,  reference epoch $T_0$, 
binary orbital period $P_b$, argument of periastron $\omega_0$,
mean advance of periastron rate $\langle\dot{\omega}\rangle$, time dilation / gravitational redshift amplitude $\gamma$, and orbital period change rate
$\dot{P}_b$.

\begin{table}

\caption{Measured Orbital  Parameters}
\begin{tabular}{lll}
\tableline
Fitted Parameter			&Value		&Uncertainty     \\
\tableline

$a_p \sin$ i (s)			&  2.341774	& 0.000001    \\
e 					&  0.6171338	& 0.0000004   \\
T$_0$  (MJD)		    	     	& 46443.99588317& 0.00000003    \\
P$_b$  (d)				& 0.322997462727& 0.000000000005 \\
$\omega_0$ (deg)			& 226.57518	& 0.00004       \\
$\langle\dot{\omega}\rangle$ (deg/yr)	& 4.226607	& 0.000007      \\
$\gamma$ (s)				& 0.004294	& 0.000001      \\
$\dot{P}_b$  ($10^{-12}$ s/s)		& -2.4211	& 0.0014       \\
\tableline
\tableline
\end{tabular}
\end{table}

The results for these eight fitted parameters, and their formal 1-$\sigma$ 
uncertainties, are listed in Table 1.  Only seven are required in order to
specify completely all the orbital parameters and component masses except
for an unknown 
rotation about the line of sight.  For example the relativistic measurables
$\langle\dot{\omega}\rangle$ and $\gamma$ depend on the Keplerian 
parameters and the
companion masses $m_p$ and $m_c$ in the following fashion:

\begin{equation}
\langle\dot{\omega}\rangle= 3 \  G^{2/3} \  c^{-2} \
(P_b / 2 \pi)^{-5/3} \  (1-e^2)^{-1} \ (m_p + m_c)^{2/3},
\end{equation}
and
\begin{equation}
\gamma= G^{2/3} \  c^{-2} \ e \
(P_b / 2 \pi)^{1/3} \ m_c \ (m_p + 2 m_c) \ (m_p + m_c)^{-4/3}.
\end{equation}
Consequently measurement of the Keplerian paramaters and 
$\langle\dot{\omega}\rangle$ and $\gamma$ permits us to solve these two
equations simultaneously for component masses, yielding
$m_p=1.4408\pm0.0003 \ M_{\sun}; \ m_c=1.3873\pm0.0003 \ M_{\sun}$, where
the uncertainty in the Newtonian Gravitational Constant $G$ is not included.
A graphical solution for the masses, also using Equations (1) and (2) and
the same measurables, is shown in Figure 1.

\begin{figure}
\plotfiddle{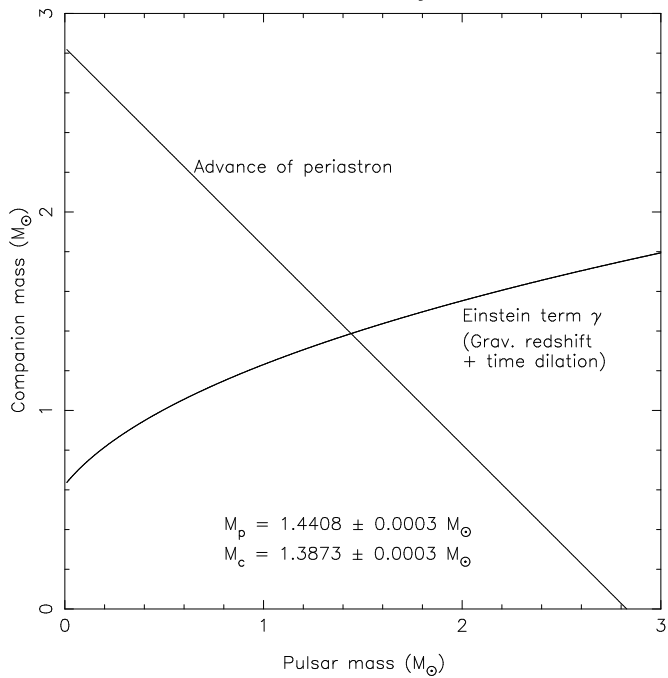}{3.0 in}{-90}{50}{50}{-200}{280}
\caption{Constraints on the pulsar and companion mass from measurements
of the ``Keplerian'' orbital elements plus relativistic advance of periastron
$\langle\dot{\omega}\rangle$ and the ``Einstein'' term $\gamma$.  The 
uncertainties in $\langle\dot{\omega}\rangle$ and  $\gamma$ are 
smaller than the displayed linewidths.}
\end{figure}

These seven measured quantities and the parameters derived from them, such as
$m_p$ and $m_c$, are then adequate to derive all other interesting
orbital parameters, such as the sine of the orbital inclination $\sin$ i and the 
various semimajor axes $a, a_p$, and $a_c$:

\begin{equation}
\sin i = G^{-1/3} \ c \ (a_p\sin i/m_c) \ (P_b/2\pi)^{-2/3} \
(m_p+m_c)^{2/3},
\end{equation}
\begin{equation}
a = G^{1/3} \ c^{-1} \ (P_b/2\pi)^{2/3} \ (m_p+m_c)^{1/3},
\end{equation}
\begin{equation}
a_p = a \ m_c \ (m_p + m_c)^{-1},
\end{equation}
and
\begin{equation}
a_c = a \ m_p \ (m_p + m_c)^{-1}.
\end{equation}

In earlier datasets we also solved for the two parameters of the
Shapiro gravitational delay, $r$ and $s$, and we were marginally able to  
constrain them (Taylor and Weisberg 1989; Taylor et al 1992).  As the orbit
 has precessed,
however, the geometry has become less favorable for measuring this phenomenon.
Continuing orbital precession is now causing the signature to begin to grow 
again.

\section{Gravitational Radiation Emission}

The emission of gravitational radiation should lead to orbital energy loss
and orbital decay.  This will be observable as an orbital period change,
$\dot{P_b}$.  According to general relativity, the orbital period change
due to gravitational radiation damping will depend on the system parameters 
as follows:
\begin{eqnarray}
\dot{P_b}=-\frac{192\ \pi \ G^{5/3} } {5 \ c^5}  \ 
\left(\frac{P_b}{2\pi}\right)^{-5/3} \  (1-e^2)^{-7/2} \   \times \\
\nonumber \left(1 + \frac{73}{24} e^2 + \frac{37}{96} e^4\right) 
 \ m_p \ m_c \ (m_p + m_c)^{-1/3}.
\end{eqnarray}

Inserting our measured values of the parameters given 
in Table 1 into Equation (7), we calculate  that the  general relativistic 
prediction for orbital period decay rate is 
$\dot{P}_{b \ GR} = (-2.40247\pm0.00002) \times 10^{-12}$ s/s.

The galactic acceleration of the system and the Sun also cause an orbital 
period change, $\dot{P}_{b, gal}$ (Damour \& Taylor 1991).   Using their
Equation (2.12) and their parameters for  $(\dot{P}_b / P_b)_{gal}$ [except for
the following updated  parameters:  solar galactocentric distance $R_0 =	
8.0\pm0.5$ kpc (Reid 1993);  pulsar distance  $d =5.90\pm0.94$ kpc  (Cordes \& 
Lazio 2002);  solar galactocentric circular velocity $v_0 = 224\pm16$ km / s 		    	     	(Reid et al 1999, Backer \& Sramek 1999);  pulsar proper motion 
$\mu = 2.6\pm0.3$ mas / yr], we find that  $\dot{P}_{b \ gal} = 
-(0.0125\pm0.0050) \times 10^{-12}$ s/s.  The galactic quantities will be 
improved in the next several years with additional interferometric observations
of Sgr A* and infrared measurements of stars orbiting it (Salim \& Gould 1999; 
Ghez et al 2000; Schodel et al 2002).

Correcting the observed orbital period derivative for the galactic acceleration term,

\begin{equation}
\dot{P}_{b \ corrected} = \dot{P}_{b \ observed} - \dot{P}_{b \ gal}, 
\end{equation}
we find $\dot{P}_{b \ corrected} = -(2.4086\pm0.0052) \times 10^{-12}$ s/s,
in good agreement with the theoretical general relativistic result 
$\dot{P}_{b \ GR}$ 
given above.  The comparison is also shown graphically in Figure 2, where
we plot the observed and general relativistically predicted 
shift of periastron time resulting from gravitational radiation damping.

This result provides strong confirmation of the existence of gravitational
radiation as predicted by general relativity.  Weisberg \& Taylor (1981)
show that it rules out numerous other relativistic theories of gravitation.
This result plus relativistic measurements of other
binary pulsars also constrain the parameters of viable tensor--scalar 
theories (Taylor et al 1992; Damour \& Esposito-Farese 2001).

\section{Geodetic Spin Precession}

The spinning, orbiting pulsar should also experience geodetic precession,
which is caused by general relativistic gravitomagnetic spin--orbit coupling 
(Damour \& Ruffini 1974; Barker \& O'Connell 1975a,b).  This will lead to
a change in the observed pulseshape
as the Earth--pulsar line of sight  cuts across different parts of the 
precessing beam.  Weisberg, Romani, \& Taylor (1989) noted the first pulseshape
changes ascribed to geodetic precession.  The pattern became clearer when
Kramer (1998) detected a secular narrowing of the separation between the 
two principal pulse components, which he was able to successfully model.
Weisberg \& Taylor (2002) measured shape changes throughout the pulse in
twenty years of Arecibo 21 cm data.  The  pulse profile was divided into
even and odd components to separate overall beam structure from patchy 
structure, respectively. We found that the outer skirts of the even
profile barely changed while the peaks drew closer and the central saddle
filled in (Figure 3a).  Our  model that best fits these data indicates an 
hourglass-shaped beam elongated in the latitude direction (Figure 3b).  
Interestingly, Link \& Epstein
(2001) found a similar beam shape for a (nonrelativistically) precessing
pulsar, B1828-11.  See also Kramer (this volume) for additional observations 
and modeling.

\begin{figure}
\plotfiddle{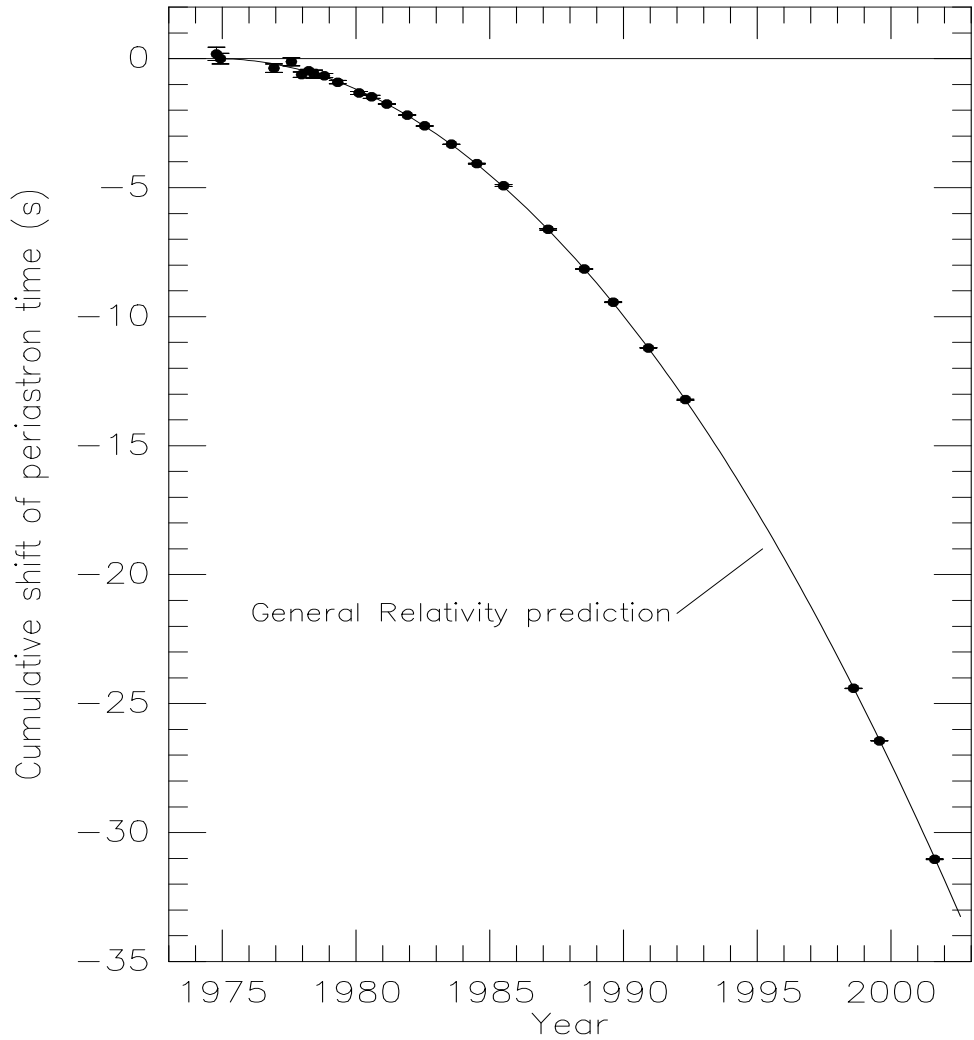}{4. in}{0}{70}{60}{-230}{-50}
\caption{Gravitational radiation damping should cause orbital decay which
leads to an accumulating shift in epoch of periastron.  The parabola 
illustrates the general relativistically predicted shift, while the 
observations are marked by data points.  In most cases (particularly in the 
later data), the measurement uncertainties are smaller than the line
widths.}
\end{figure}

\begin{figure}
\plottwo{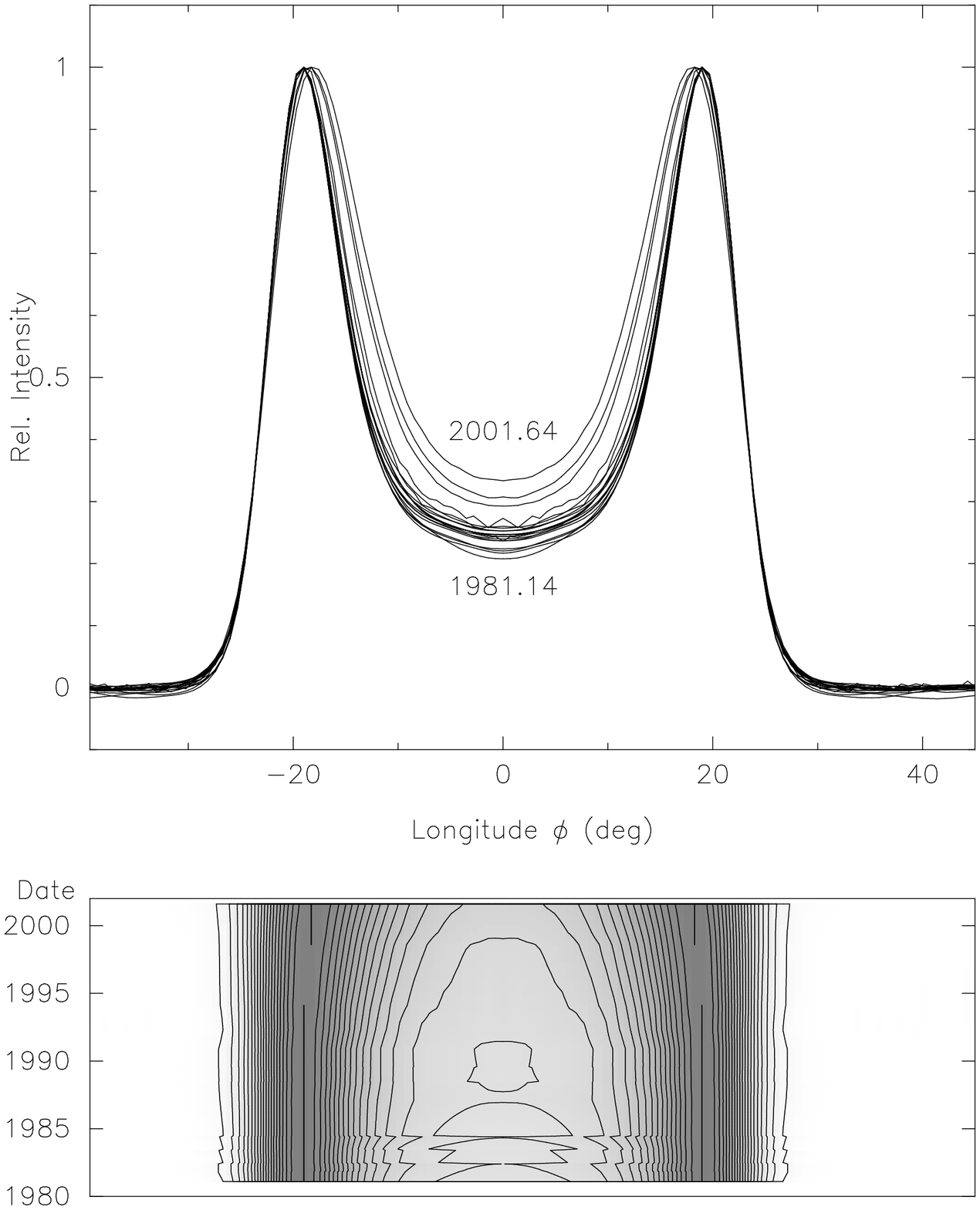}{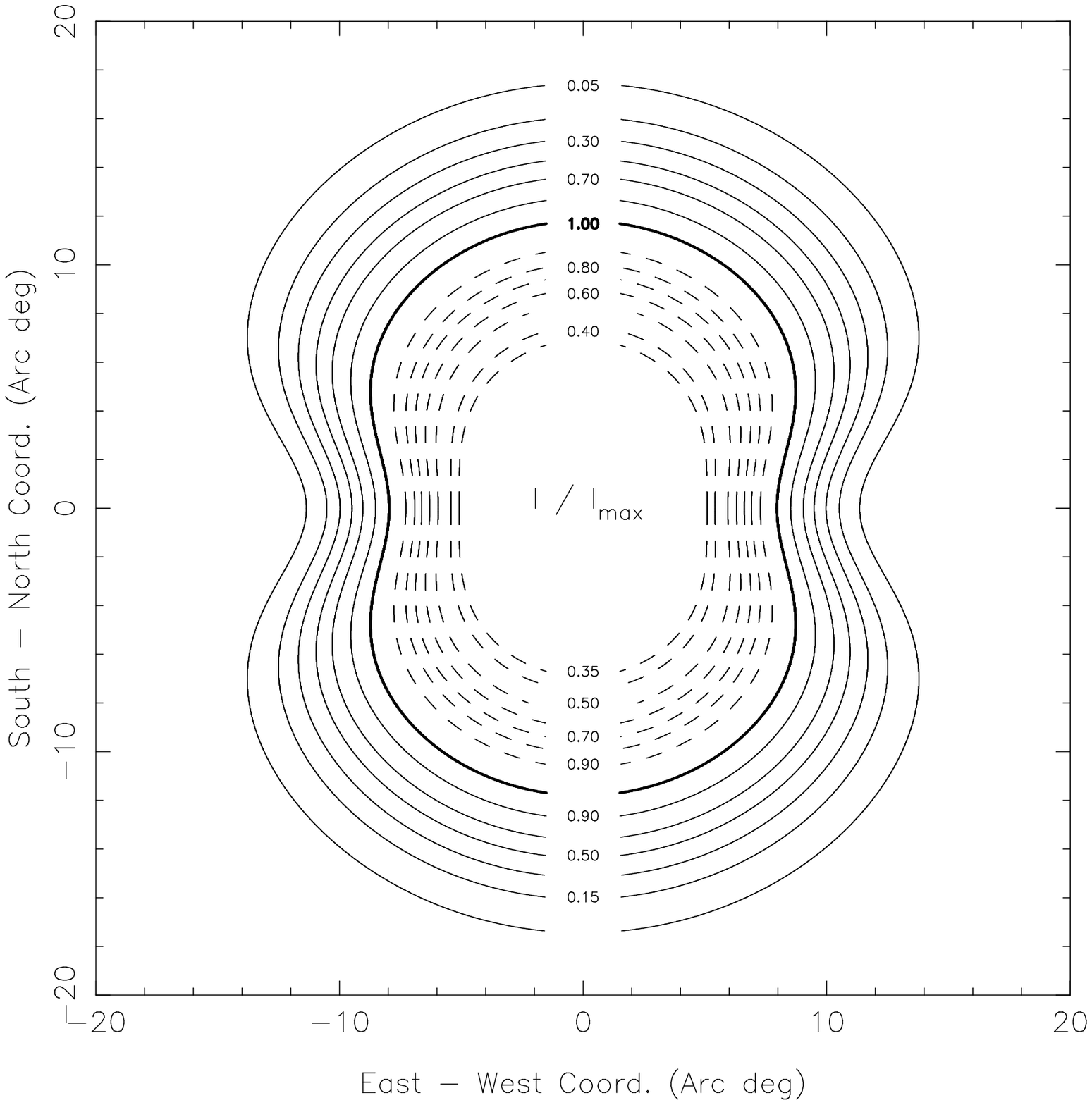}
\caption{Left: Even component of B1913+16 pulse profile at 21 cm versus date.  
Right:  Our inferred best--fitting emission beam model.}
\end{figure}

\acknowledgements{We thank the National Science Foundation for supporting this
project since its inception (most recently through grants AST 00-98540 and 
96-18357), and the staff of Arecibo Observatory for their
enthusiastic help.  Arecibo Observatory is operated by Cornell University
under cooperative agreement with the NSF.}

\end{document}